# Realization and Properties of Biochemical-Computing Biocatalytic XOR Gate Based on Enzyme Inhibition by a Substrate

Jan Halámek, Vera Bocharova, Mary A. Arugula, Guinevere Strack,
Vladimir Privman and Evgeny Katz*

*Department of Chemistry and Biomolecular Science, and*
*Department of Physics, Clarkson University, Potsdam, NY 13699*

**\*** E-mail: ekatz@clarkson.edu

## ABSTRACT

We consider a realization of the **XOR** logic gate in a process biocatalyzed by an enzyme which can be inhibited by a substrate when the latter is inputted at large enough concentrations. A model is developed for describing such systems in an approach suitable for evaluation of the analog noise amplification properties of the gate. The obtained data are fitted for gate quality evaluation within the developed model, and we discuss aspects of devising **XOR** gates for functioning in "biocomputing" systems utilizing biomolecules for information processing.

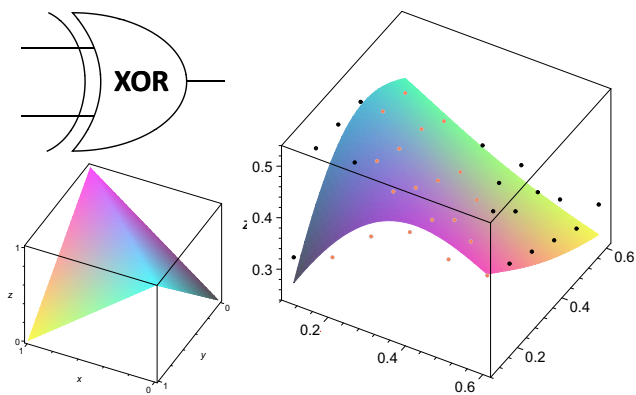

Realization and model analysis of **XOR** biomolecular logic based on enzyme, HRP, biocatalyitc function being inhibited at elevated concentrations of its substrate, $H_2O_2$.

1.    Introduction

Chemical computing,[1-7] i.e., processing of information with chemical reactions, has been an active area of unconventional computing.[8,9] Several chemical systems carrying out Boolean logic gates, such as **AND**,[10,11] **OR**,[12] **NAND**,[13,14] **NOR**,[15-18] **INHIB**,[19-22] **XOR**,[23-26] have been demonstrated. Biochemical computing[27-29] has utilized biomolecular kinetics for information processing, realizing binary logic gate functions and few-gate networks based, for instance, on proteins/enzymes,[30-38] antigens/antibodies,[39,40] DNAzymes,[41,42] DNA,[43,44] RNA[45-49] and whole cells.[50] Biocomputing systems offer approaches to design multi-signal responsive biosensors[51-53] and bioactuators[54,55] processing complex patterns of biochemical signals, for potential biomedical applications.[56-59]

**XOR** (e**X**cluded **OR**) gate has been a binary logic operation most difficult for (bio)chemical realizations.[60] Indeed, **XOR** is expected to yield output **0** when the two inputs are applied at zero levels (input combination **0,0**). When only a single input signal is applied at level **1**, output **1** is expected (system "activation" for input combinations **0,1** and **1,0**). However, simultaneously applying both inputs at level **1**, keeps the system inactive: output **0** (for inputs **1,1**). Therefore, unlike other recently realized[10-22] chemical gates (**AND**, **OR**, **NAND**, **NOR**, **INHIB**), the **XO**R function involves an inherently non-monotonic chemical output in response to increasing concentrations of the two chemical inputs.[60] In (bio)chemical setting this requires reactions with kinetics such that the effects of two input species cancel each other when they are both present in the system.

The **XOR** function has been realized by mutual negation of the chemical inputs which drive molecular restructuring, such as acid and base.[23,24] In supra-molecular systems input signals resulted in a change of a chemical complex's structure, measured by changes in optical absorbance or fluorescence, as each input binds at different sites of the complex, whereas binding of both of them resulted in no changes in the optical properties.[25,26] Biomolecular realization can be based on biorecognition or biocatalytic properties of DNA[48] or enzymes.[60-64] In "signal change" enzyme-based **XOR** gates[38,60-63] oppositely directed biocatalytic reactions are driven by input signals. Separate application of each chemical input unbalances the system, resulting in the output signal as a change in a measurable property. Both inputs, however, when present, drive competing reactions thus keeping the balance (no signal change). The resulting optical[60-63] or electrochemical outputs[38] require certain post-processing in order to get the absolute value of the change as the **XOR** value. Such gates have been used as model systems[60] for **XOR** realizations. Here is considered a different approach whereby the non-monotonic response is achieved by a suppression mechanism of the oxidative function of the enzyme HRP (horseradish peroxidase) at increased concentrations of its primary substrate, $H_2O_2$. The experimental details are given in Section 2.

Despite the difficulties[60] in realizing the **XOR** gate in terms of single- or few-step (bio)chemical processes, explorations of its properties are important for several reasons. This gate is among the basic computing elements of half-adder/half-subtractor[65-68] or full-adder/full-subtractor.[69,70] These arithmetic functions were realized using DNA-based[71] and enzyme-based[64] processes. **XOR** is also a part of reversible gates: **CNOT**, **CCNOT**, etc., that have been utilized in other unconventional computing realizations.[72-77]

Furthermore, a response-surface function of a system that yields **XOR** output at the logic-point inputs **00**, **01**, **10**, **11**, has a convoluted, saddle shape[60] as compared to other gates, such as **AND**. This matter is addressed in Section 3, which also introduces a model that allows analysis of the noise-handling properties of our specific **XOR** realization. Generally, we expect[60] and confirm that the **XOR** gate is more noisy that **AND**, **OR** and other gates with less "structured" response surfaces. While the study of the noise handling by **XOR** gates is interesting, such gates will be candidates for inclusion in larger networks only provided noise suppressing elements (filters) are utilized, rather than being directly optimizable. Section 4 presents model fit and analysis of our data, discussion of the results, and offers concluding remarks.

## 2. Experimental: XOR Gate Based on Enzyme Suppression by Substrate

*Chemicals and Materials*

Glucose oxidase (GOx) from *Aspergillus niger* type X-S (EC 1.1.3.4), horseradish peroxidase (HRP) type VI (EC 1.11.1.7), potassium hexacyanoferrate (II) trihydrate and D-(+)-glucose (99.5%) were purchased from Sigma-Aldrich and used without further purification.

H$_2$O$_2$ (30% w/w) was purchased from Fisher. Ultrapure water (18.2 MΩ·cm) from NANOpure Diamond (Barnstead) source was used in all of the experiments.

*Composition and mapping of the **XOR** System*

The main process of the **XOR** gate was based on 0.1 µM of enzyme HRP catalyzing the oxidation of 1 mM ferrocyanide by the "input" H$_2$O$_2$, in 10 mM phosphate buffer, pH 7.4. Glucose, pre-processed to yield the oxidizer: H$_2$O$_2$ (as described below and further explained in Section 3) and additional directly introduced H$_2$O$_2$ were used as the two input signals. For both inputs, the logic value **0** was chosen as 0.1 mM and logic value **1**, as 0.6 mM.

For each experiment, Input 1, glucose, was pre-processed in a separate cuvette by incubation with the 4.76 U·mL$^{-1}$ of GOx enzyme. The enzymatic reaction was continued for 15 min in the total volume of 2 mL with the constant air bubbling through the reaction solution. After that, 1 mL of the reaction mixture was transferred to another cuvette, where it was mixed with the Input 2. Input 2, H$_2$O$_2$, was directly added from the 100 mM stock by adjusting volume of transferred H$_2$O$_2$ solution according to the needs of experiment. Premixed solution of the inputs was transferred to an optical cuvette already containing HRP and ferrocyanide. The schematic illustration of the experimental set-up is represented in Figure 1.

In the gate-response-mapping experiments the input signals were applied at (separately) variable concentrations, glucose: 0.1, 0.2, 0.3, 0.4, 0.5 and 0.6 mM, and H$_2$O$_2$: 0.1, 0.2, 0.3, 0.4, 0.5 and 0.6 mM. There were thus 36 experimental data sets.

*Optical Measurements*

Absorbance, A, measurements were performed using a UV-2401PC/2501PC UV-visible spectrophotometer (Shimadzu, Tokyo, Japan) at (25.0 ± 0.2) °C. The reaction was carried out in a 1 mL poly(methyl methacrylate), PMMA, cuvette for all the different input combinations. The increase of absorbance was monitored as a function of time, at λ = 420 nm, resulting from the formation of ferricyanide due to oxidation biocatalyzed by HRP. A selection of our absorbance data is shown in Figure 2. The gate time, $t_g$ = 60 sec, was selected to obtain a well-defined **XOR** function, as further discussed in the following sections. In particular, for this time, input combinations **0,0** and **1,1** give similar outputs which yields a good quality **XOR** function; see Figure 2.

Since some other chemicals can contribute in a limited extent, to the absorbance at $\lambda = 420$ nm, we used the *change* in the absorbance ($\Delta A$) as compared to its value at time zero of the HRP step of the experiment, as the optical output measure. This in turns can be converted to the chemical output signal concentration, that of ferricyanide, denoted by $F(t)$, where $t$ indicates its time-dependence, by using the Lambert-Beer law, where the molar extinction coefficient[78] of ferricyanide at $\lambda = 420$ nm is $\varepsilon_{420} = 1$ mM$^{-1}$·cm$^{-1}$. The HRP concentration (0.1 µM) was also verified spectrophotometrically, where the molar extinction coefficient[79,80] of HRP at $\lambda = 403$ nm is $\varepsilon_{403} = 102$ mM$^{-1}$·cm$^{-1}$. As explained in Section 4, we used the experimental mean values of the logic-point **0** and **1** of the output at the gate time (60 sec), at the appropriate pairs of the "logic" inputs, to define $F_{\min} = 0.32 \,\text{mM}, F_{\max} = 0.41 \,\text{mM}$, used in shifting and scaling the output $F(t_g)$ to the logic range which ideally (if there were no noise in the data) should be between numerical 0 and 1, to draw the bar chart in Figure 2, which illustrates the degree of accuracy of the realized of the **XOR** function (see the next two sections for further discussion and references).

## 3. Description and Modeling of the XOR Gate Realization

As mentioned in Section 1, among the binary logic gates realized in the framework of enzyme-catalyzed biochemical processes, **XOR** has been rather problematic, as detailed shortly. This puts in question the **XOR** gate's utility as a useful network element for biochemical computing designs. However, experimental realizations of **XOR** gates and their analysis are of importance for general interest, as well as precisely for the purpose of cataloguing the challenges associated with design and adjustability of various component biochemical processes required to achieve complex information processing. The difficulty with the **XOR** gate (bio)chemical realizations has been traced to its truth table, shown below:

| Input 1 | Input 2 | Output |
|---------|---------|--------|
| **0** | **0** | **0** |
| **0** | **1** | **1** |
| **1** | **0** | **1** |
| **1** | **1** | **0** |

(1)

Indeed, most chemical and biochemical processes yield output which is a monotonically increasing function of the input chemical concentrations. Therefore, the fact that **XOR** requires output **0** at the (**1**,**1**) inputs, necessitates the use of reactions with unusual properties or additional output-signal processing steps. Furthermore, such gates with two output values at **0** and diagonally positioned output values at **1**, have a more "structured" response surface, with therefore generally larger slopes (faster variation) than gates such as **AND**, the latter with the truth table containing only a single output **1**. As a result, the realized **XOR**-type gates typically have stronger analog noise amplification[60] and are much more difficult to optimize for noiseless network functioning than **AND**-type gates.

In our earlier work,[60] an **XOR** gate was realized based on signal change, and its noise handling properties were investigated. That realization has required certain post-processing, which was devised,[60] but not realized experimentally as a (bio)chemical step, in order to obtain the actual **XOR** response shown in (1). Furthermore, its tuning for noise reduction also proved not straightforward. Indeed, after the signal post-processing the "optimal" response surface of the signal-change gates approximates the one shown in Figure 3(a), which still has $\sqrt{2} \approx 140\%$ noise amplification factor per information processing step.[60] Here we explore another approach which yields response surfaces which are smooth functions that can approximate the ridged shape shown in Figure 3(b), which has the same noise amplification factor (if accurately realized). In fact, it is unlikely that simple biochemical steps based on properties and functioning of one or few enzymes, can approximate the planar-face "origami" shape such as in Figure 3(c), which would allow practically no noise amplification: all the faces shown have slopes 1. An even better shape is shown in Figure 3(d): a sigmoidal "filter" version of the shape in Figure 3(c), which has slopes less than 1 (means, analog noise suppression[81,82]) at all the logic points. This "saddle" shape for the **XOR** function would likely require a sequence of well-designed and tuned reaction processes of the complexity presently not realized in the (bio)chemical logic-gate literature.

In the present work, we realize the **XOR** function as shown in Figures 1 and 2. The enzyme, HRP, is known to be inhibited by too much of its substrate, $H_2O_2$. Therefore, we combine the two chemical signals, determined by the input chemicals of concentrations $I_1$ and $I_2$, into a single signal determined by $[H_2O_2]$, which in turn is driving the oxidation of a chromagen used to optically detect the output, the latter process catalyzed by HRP. As in the earlier, different realization,[60] the present **XOR** function is not carried out by a one-shot enzymatic process. Instead of the post-processing, here we have pre-processing of the inputs. Actually, we pre-processed only one input, glucose, which was fed into a system producing $H_2O_2$, catalyzed by another enzyme, glucose oxidase (GOx); see Figure 1. The other input was taken directly as $H_2O_2$. Spatial separation and temporal "clocking" of biocomputing steps has received limited attention in the literature,[83] even though it has been realized[84] that these features are important for large-scale networking of biocomputing processes. However, it is also important to note that we are interested in information processing rather than the chemical kinetics aspects of these steps. Therefore, our pre-processing was carried out for long enough (fixed) time to ensure that all glucose is processed and equal quantity of $H_2O_2$ is produced. In terms of the concentrations of glucose and $H_2O_2$, the input $I_1$ is the same.

However, we note that such pre-processing and other "black box" (means, carried out identically without regard to the actual value of the input) information processing steps can be sources of *noise*, which is our primary topic of concern in this work. Therefore, while seemingly chemically trivial, such steps should be included if we want a realistic experiment as far as data nosiness is concerned. We chose one such input pre-processing step for the present work: Glucose $\to$ $H_2O_2$; see Figure 1. The resulting data noisiness is illustrated in Figure 4. The actual kinetics is not interesting as far as information processing goes, and, as mentioned, we took the conditions such that the input, $I = [H_2O_2]$, into the second processing step, catalyzed by HRP, was $I = I_1 + I_2$.

We will use the following simple "cartoon" kinetic model to consider the parameter selections for the **XOR** function. As emphasized earlier,[27,60,81,82,85] analysis of the shape of response surfaces is frequently done with limited data and is aimed at an approximate global description of the function shape within a few-parameter model. Therefore, here we focus on two key features of the processes involved in the HRP functioning. We assume that the steps[86] of the enzymatic process which actually result in the oxidation of the chromagen, in out regime can be represented by a single-rate irreversible production rate, $RE(t)$, which can be controlled by (is approximately proportional to) the instantaneous enzyme concentration. Approximately, we thus write the first term in

$$\frac{dI(t)}{dt} = -R\,E(t)I(t) - rI^2(t)E(t) = -[R + rI(t)]I(t)E(t)\,. \qquad (2)$$

We then assume (the second term) that the inhibition by $H_2O_2$, is approximately represented by that, some of the enzyme is "taken out of the game" by $H_2O_2$, at the rate $rI^2(t)$:

$$\frac{dE(t)}{dt} = -rI^2(t)E(t)\,. \qquad (3)$$

The latter process requires another molecule of $H_2O_2$. The quadratic in $[H_2O_2]$ rate expression is obviously an oversimplification of the actual inhibition process which involves[87,88] in part an intermediate complex of HRP (with $H_2O_2$) in the enzymatic cycle, being converted to an inert species as it further reacts with $H_2O_2$, and some other secondary and additional processes.[87,88] The proposed rate expression treats these processes in a cavalier way. The fact that ferrocyanide also competes for the same main complex is also not carefully considered. Technically, this should be reflected in that the rate $r$ will depend on (decrease with) the initial ferrocyanide concentration.

The main virtue of the present "cartoon" model is that, it has few parameters (two rates). Another useful feature is that, at least in principle the model offers a prescription for how to change the rates: $R$ – by varying the initial enzyme concentration to which this rate constant should be roughly proportional; $r$ – by changing the initial ferrocyanide concentration, the increase of which will result in the effective value of $r$ decreasing. We also note that the main effects of the inhibition occur during the initial time scales of the reaction when the concentration of $H_2O_2$ is not yet significantly depleted. Therefore, for purposes of the actual calculations we can further approximate the quadratic term by replacing

$$I^2(t) \to I(0)I(t). \tag{4}$$

This makes the model analytically solvable, to yield, from

$$\frac{dF(t)}{dt} = 2RE(t)I(t), \tag{5}$$

after a detailed calculation, the expression for the concentration, $F(t)$, of the produced ferricyanide,

$$F(t) = \frac{2I_0 RE_0[1 - e^{rI_0^2 t - (R+rI_0)E_0 t}]}{(R+rI_0)E_0 - rI_0^2 e^{rI_0^2 t - (R+rI_0)E_0 t}}, \tag{6}$$

where (5) is, within the present model, the rate of its production in the peroxidase cycle.[86] Here the time, $t$, is that of the HRP part of the experiment, and for brevity we denoted the initial concentrations $I_0 = I(0)$ and $E_0 = E(0)$.

The model equation (6), fits the time dependence of the measured data, $F(t)$, for each of the set of values of the initial inputs $I_{1,2}(0)$ with $E_0 = 10^{-4}$ mM and $I_0 = I_1(0) + I_2(0)$. However, we get large variations ($\pm 80\%$) in the effective values of the rates, which is an indication that the model is at best semi-quantitative. Nevertheless, the model allows to illustrate the issues related to the selection of the XOR response shape from the nonmonotonic biochemical-process signal profile, as well as to estimate the noisiness of the resulting gate function. Indeed, Figure 5 shows sample curves of the dependence of $F(t)$ on $I_0$ for a typical set of fitted parameters, $R = 208$ mM$^{-1}$, $r = 0.133$ mM$^{-2}$, for several fixed times. It is obvious that the logic zero for the output cannot be at the physical zero. However, in biomedical applications[53,58,59,89,90] this is not a disadvantage because both normal and pathophysiological concentrations of most biochemicals are typically non-zero.

If we select some reference value, $I_{1,2}(0) = I_{\min}$, as logic-**0** of the inputs, and fix the gate-time, $t = t_g$ of the logic-gate processing, then for an accurate **XOR**, the value of $F_{\min} = F(t = t_g; I_0 = 2I_{\min})$ should also be obtainable at a larger, logic-**1** (for both inputs) $I_{1,2}(0) = I_{max}$, i.e., $F_{min} = F(t = t_g; I_0 = 2I_{max})$. This is obviously possible, because the curves in Figure 5 are non-monotonic, as long as the selected $I_{min}$ is not too small. However, then the ligic-**1** value of the output, $F_{max} = F(t = t_g; I_0 = I_{min} + I_{max})$, will be at a certain large value but is not guaranteed to be close to the maximum of the curve for the selected gate time. The result will be that the desired shape of Figure 3(b), will not be well approximated as a "smoothed ridge." The response surface, which depicts the dependence of the scaled and shifted output variable

$$z = \frac{F - F_{min}}{F_{max} - F_{min}}, \qquad (7)$$

on the similarly scaled and shifted inputs,

$$x = \frac{I_1(0) - I_{min}}{I_{max} - I_{min}}, \qquad y = \frac{I_2(0) - I_{min}}{I_{max} - I_{min}}, \qquad (8)$$

will instead be asymmetrical and bulging above the level of the logic-1 for *z*, for certain inputs' (*x* and *y*) values between 0 and 1. The result will be large slopes and thus large analog-noise amplification for the this type of gates. This is precisely the pattern of behavior found in our data, as described in the next section. Thus it is complicated/impractical to "optimize" by parameter selection/modification, the functioning of this type of **XOR** gates to achieve a smaller amplification factor (really close to the optimal ~ 140%) for analog noise in the inputs.

## 4. Results and Discussion

Our data were taken as time-dependent sets measured for several values of both inputs. The data generally follow the pattern predicted by the semi-quantative model of Section 3. An "optimization" by adjusting the parameters choices to get a response surface from which an **XOR** shape can be "cut out" which is not too asymmetrical and bulging as compared to Figure 3(b), suggested the value $E(0) = 10^{-4}$ mM for the initial enzyme concentration used, as well as the choices of $I_{min} = 0.1$ mM, $I_{max} = 0.6$ mM, with the gate-time for the HRP processing step $t_g = 60$ sec; see Figure 2. The measured data have some noise and also do not exactly follow the analytical expression, (6), of our semi-quantitative model.

The response surface at the selected gate-time was then mapped out based on the data for $I_{1,2} = 0.1, 0.2, 0.3, 0.4, 0.5, 0.6$ mM (total 36 data points). This gave the data depicted in Figure 6, which also shows the fitted surface according to the model equation (6). The model fit is semi-quantitative, and here yield estimates $R = (208 \pm 14)\,\text{mM}^{-1}$, $r = (0.133 \pm 0.013)\,\text{mM}^{-2}$ However, as emphasized in Section 3, much larger variation of the rates is obtained if we consider separate time-dependent data sets.

Since the data are noisy and fluctuate about the fitted surface (see Figure 6), as mentioned in Section 2 we used the experimental mean values of the logic-point **0** and **1** for the output, at the "logic" inputs, to define $F_{min} = 0.32\,\text{mM}$, $F_{max} = 0.41\,\text{mM}$, used in shifting and scaling the output, according to (7); see the bar chart in Figure 2. The values in this chart, and those in Figure 6 illustrate that there are several sources of noise in single-gate functioning. Specifically, the "logic" values of the output are not exactly at the desired numerical 0 and 1. Furthermore, the data are generally noisy, so that various experimental realizations do not even reproduce a smooth scaled surface *z*(*x*,*y*), but rather a certain distribution of the *z*-values. However, as emphasized in earlier works,[27,60,81,82,85] for *network functioning* of the gate as part of complex information processing designs, *noise amplification* in the gate's signal processing is frequently more important than the noise-to-signal relative magnitude.

In order to estimate the analog noise amplification factor[27,60,81,82,85] assuming small spread of the noise in the inputs, we can use the absolute values of the gradients of the response surface function, *z*(*x*,*y*), at the logic points. These gradient values were calculated for our fitted model surface, yielding 11.2, 0.6, 2.4, 2.4, for the logic inputs **00**, **11**, **10**, **01**, respectively. We point out that had the data accurately approximated the "ridged" shape in Figure 3(b), we would approach the optimal (for this type of gate) noise-amplification factor of $\sqrt{2} \approx 140\%$. However, we get (the largest, at **00**) factor of eleven-fold amplification, $11.2 \approx 1100\%$. Thus, the realization is not optimal, and unfortunately, is not easy to improve in any significant way because of the numerous restrictions imposed on the parameter selection by the requirement of approximating the **XOR** truth table, (1), as described here and in Section 3.

In conclusion, in this work we experimentally realized and theoretically modeled a new **XOR** function response, of the type shown in Figure 3(b), for enzyme-based logic, which has the advantage of not requiring the post-processing which is needed and was contemplated, but not yet experimentally realized for the earlier reported "signal change" **XOR**s approximating response of the type shown in Figure 3(a). As with all the biochemical **XOR** functions based on single or few reaction steps, the realized gate function based on a substrate at larger concentrations inhibiting the enzyme's function, is noisy and, unlike **AND** or gates similar to it, cannot be as easily optimized at the biochemical reaction parameter level. Additional optimization by filtering and network design, as discussed elsewhere,[91,92] is needed for scalable functioning for complex information processing.

The authors acknowledge research funding by the NSF (grant CCF-1015983).

**Figures**

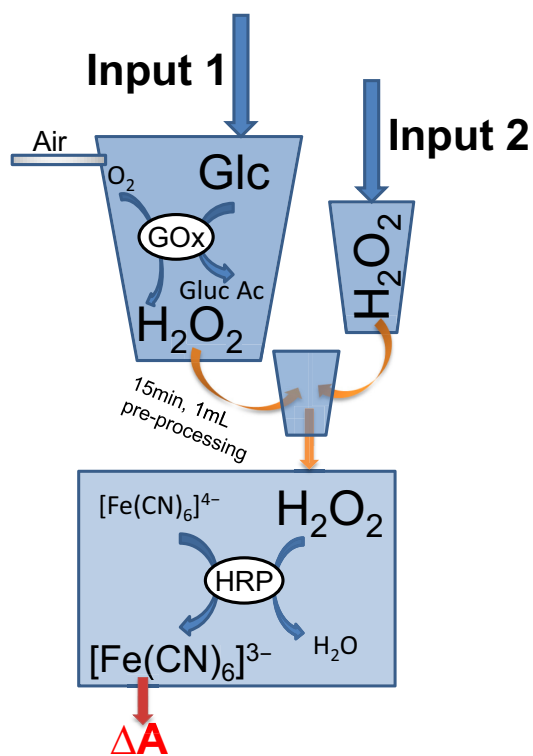

**Figure 1.** Biocatalytic cascade initiated by Input 1, glucose (here marked as Glc), and Input 2, $H_2O_2$, with the level of ferricyanide measured as the output signal of the **XOR** function. For this design Input 1 was pre-processed with GOx in the cuvette of 2 mL with the air bubbled through the reaction solution. After 15 min of this enzymatic reaction in which $H_2O_2$ and gluconic acid (here marked as Gluc Ac) were produced, 1 mL of the resulting solution was transferred to another cuvette, where was mixed with Input 2. The premixed solution was then transferred to an optical cuvette in which ferrocyanide to oxidized by $H_2O_2$, by the biocatalytic action of HRP, to yield ferricyanide. The kinetics of formation of ferricyanide was optically monitored at $\lambda = 420$ nm.

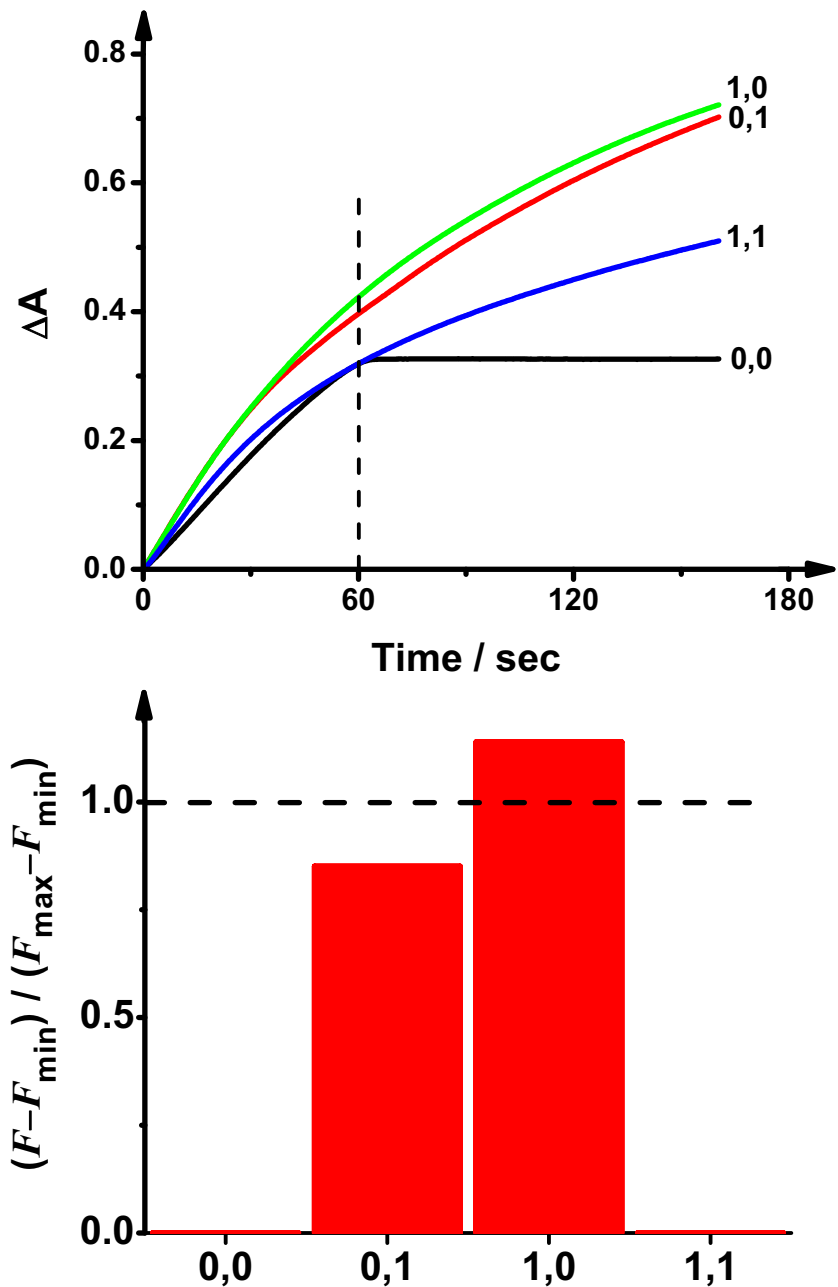

**Figure 2.** Time-dependent optical signals (top panel) measured as the change is the absorbance due to the production of ferricyanide, for the four "logic value" combinations of the two inputs: glucose and $H_2O_2$. For both inputs, logic **0** is taken at 0.1 mM and logic **1**, at 0.6 mM. Optical absorbance measurements were performed at $\lambda$ = 420 nm. The histogram (bottom panel) features the **XOR** logic realization at the selected gate time of 60 sec. The vertical axis gives the ferricyanide concentration shifted and scaled in term of the logic-value range between values selected as **0** and **1** for the output, as explained in the text.

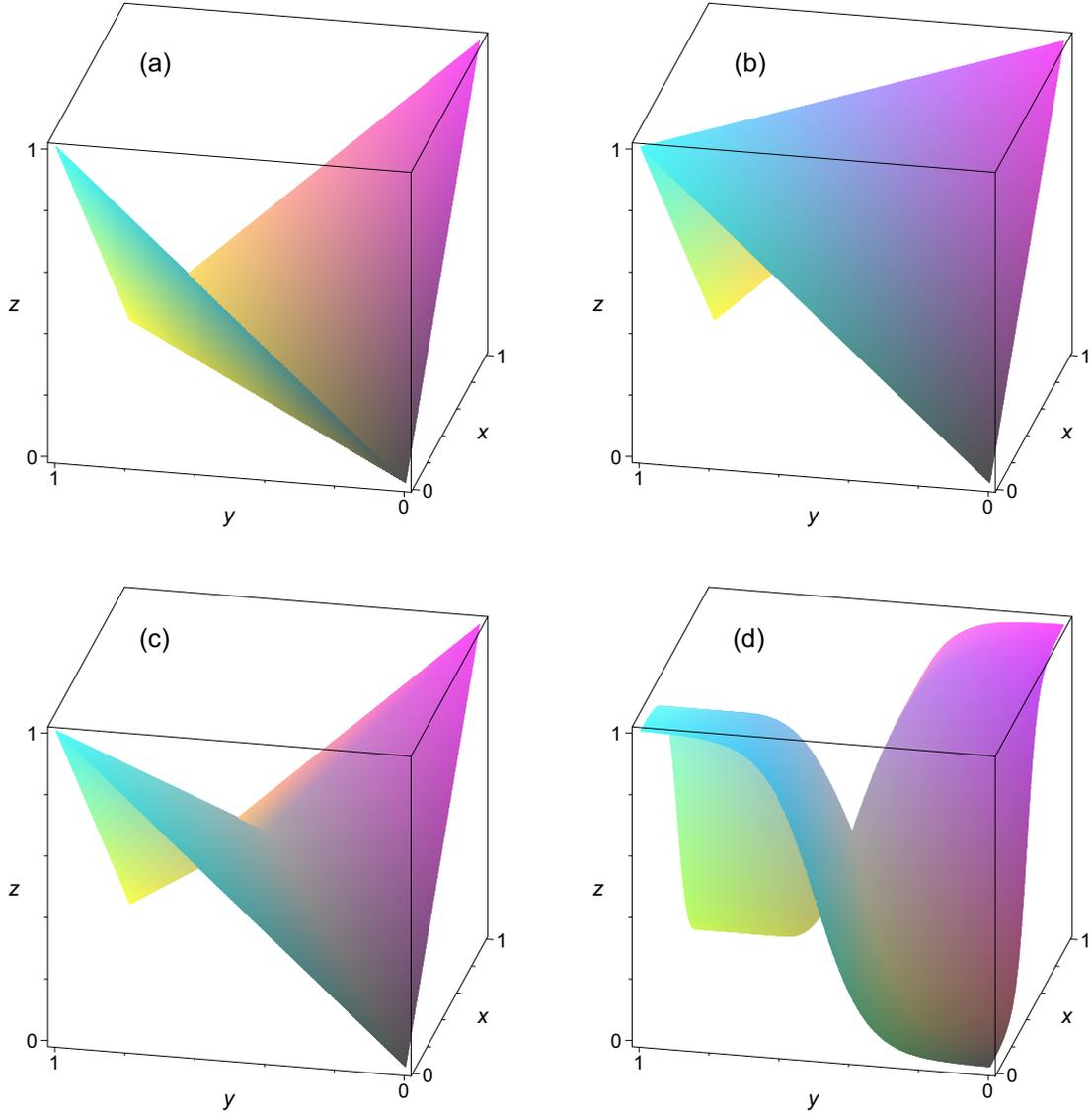

**Figure 3.** Response functions of interest in the **XOR** gate realizations. All the surfaces shown realize the **XOR** truth table, see (1), as the $z$-values of the function $z(x,y)$ at the four logic points $(x,y) = (0,0), (1,0), (0,1), (1,1)$. Experimentally realized: (a) the shown ridged shape can be approximately realized by the earlier-reported[60] signal-change **XOR** function, and has the optimal analog noise amplification factor, $\sqrt{2} > 1$, for such gates; (b) the shown ridged shape can be approximated by the **XOR** process realized in the present work. Not realized, and likely not realizable with single- or few-enzyme reactions: (c) an "origami" planar shape without noise amplification; (d) a "saddle" shape which would be optimal for noise-suppressing **XOR** gates.

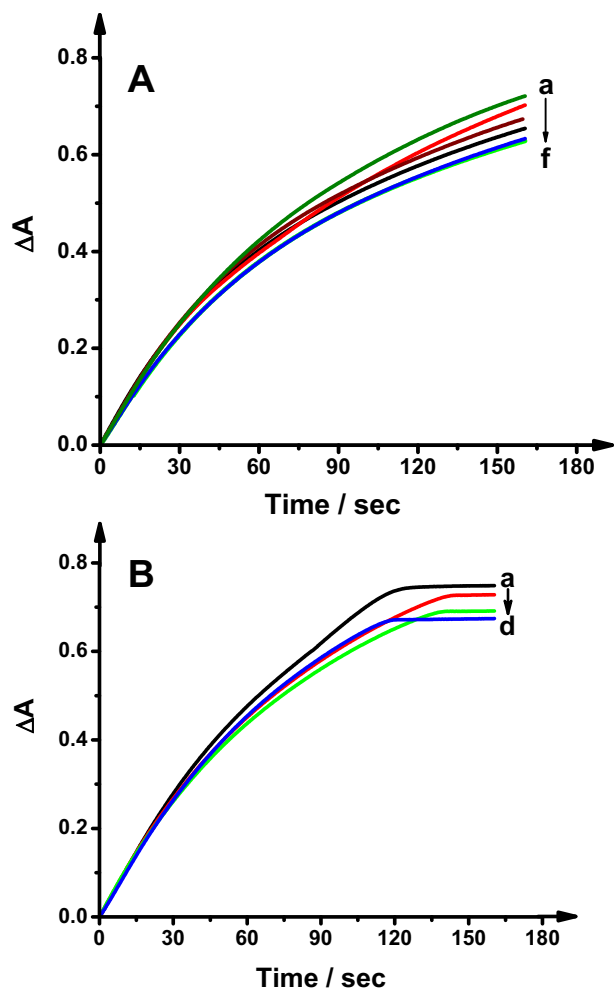

**Figure 4.** Data sets obtained for several combination so the inputs which, after the pre-processing of glucose should yield an identical overall $H_2O_2$ input into the second, HRP stage of the reaction (see Figure 1) and therefore, result in identical outputs. However, nosiness of the data results (in part) due to a different fraction of the overall $H_2O_2$ input fed directly vs. that produced from the glucose-involving process by the action of GOx. Panel A presents kinetic (time-dependent) data for inputs: (a) glucose 0.6 mM and $H_2O_2$ 0.1 mM; (b) glucose 0.1 mM and $H_2O_2$ 0.6 mM; (c) glucose 0.5 mM and $H_2O_2$ 0.2 mM; (d) glucose 0.2 mM and $H_2O_2$ 0.5 mM; (e) glucose 0.4 mM and $H_2O_2$ 0.3 mM; and (f) glucose 0.3 mM and $H_2O_2$ 0.4 mM. Panel B presents data for: (a) glucose 0.1 mM and $H_2O_2$ 0.4 mM; (b) glucose 0.4 mM and $H_2O_2$ 0.1 mM; (c) glucose 0.2 mM and $H_2O_2$ 0.3 mM; and (d) glucose 0.3 mM and $H_2O_2$ 0.2 mM. Panel B illustrates that noise is present even in the saturation regime which is reached once the available $H_2O_2$ is used up, highlighting the property that the observed noise is at least in part due to different availability of $H_2O_2$ after the pre-processing in the GOx step.

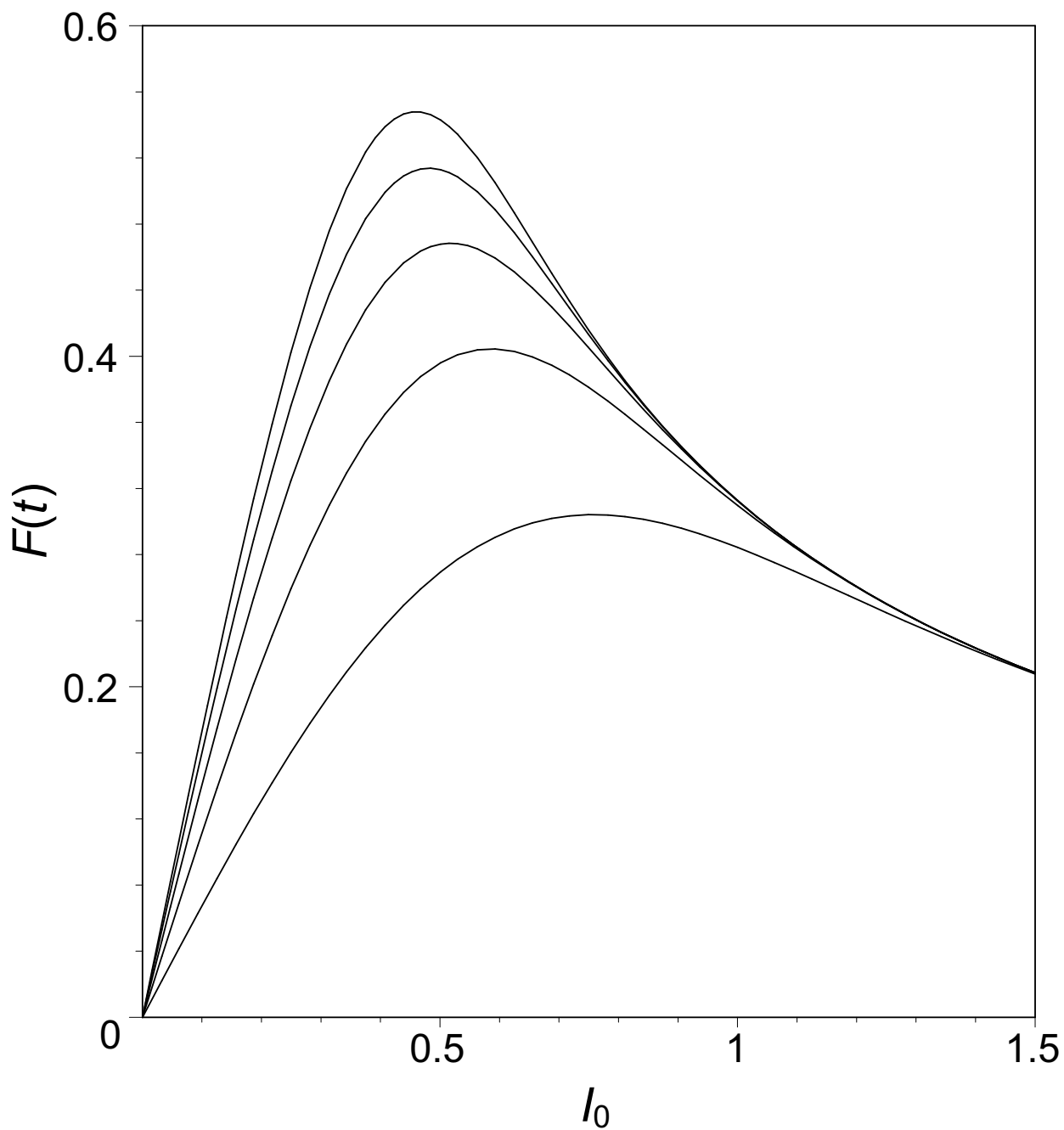

**Figure 5.** Plots of $F(t)$ as a function of $I_0$, for typical parameters reflecting our experimental conditions (see text), for several gate times $t = t_g = $ 20, 40, 60, 80, 100 sec, where increasing times give curves with larger values (of the output signal).

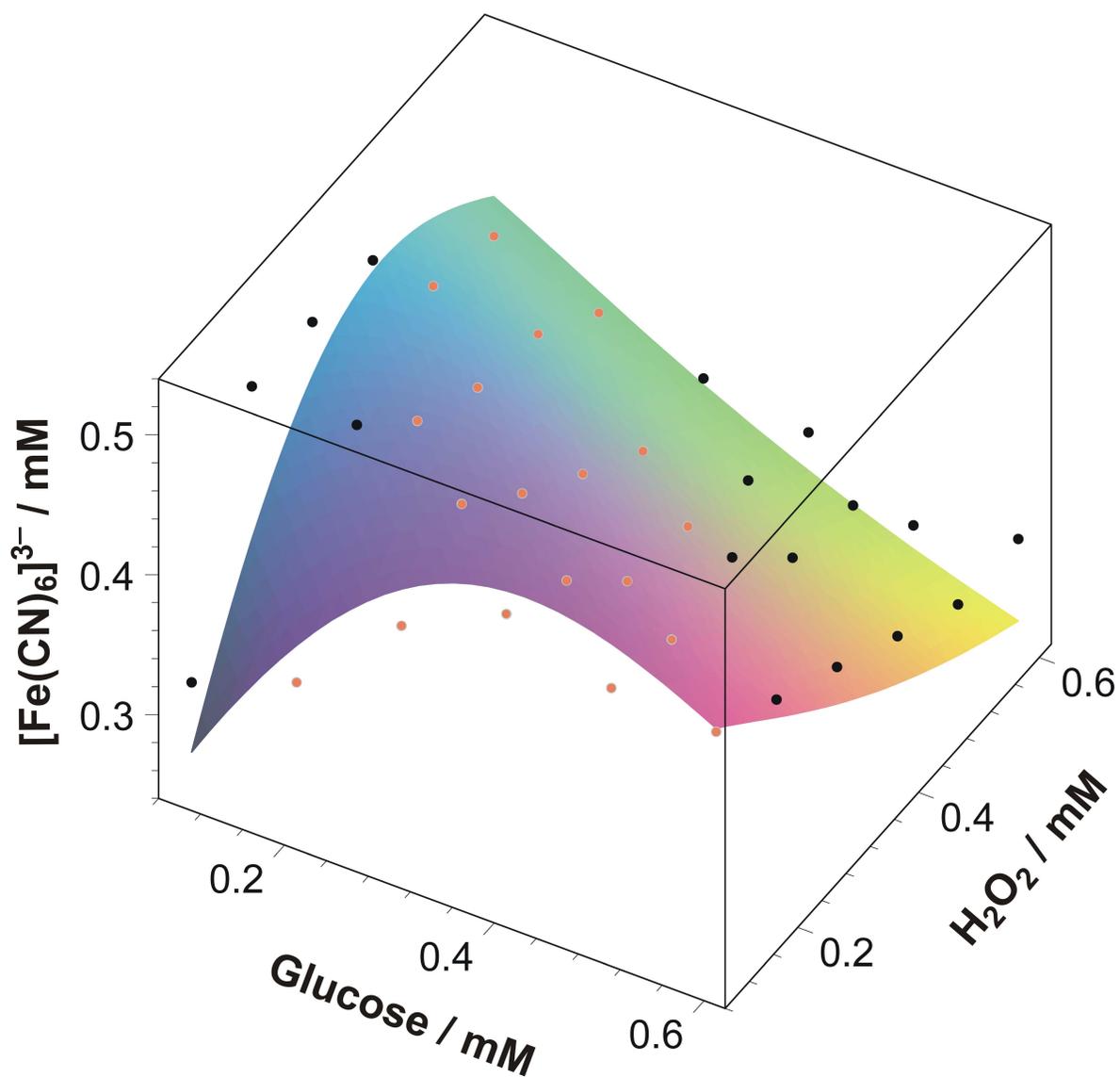

**Figure 6.** Model fit (surface) of the experimental data for the gate-function response shape (shown in terms of the physical, rather than scaled/shifted "logic" variables), at the selected gate time, 60 sec. The black circles indicate the 17 data points which are above the level of the fitted surface. The orange circles mark the 19 data points which are below the level of the fitted surface. The data points have their horizontal-plain coordinates on a grid (not shown) defined by the 36 pairs of values at concentratinos 0.1, 0.2, 0.3, 0.4, 0,5, 0.6 mM along both the Glucose and $H_2O_2$ axes.